\relax 
\documentstyle [art12]{article} 
\begin {document} 
\bibliographystyle {plain} 
%\tableofcontents
\title{\bf Capacitance and Transport through an Isolated Impurity Center in
One-Dimensional Luttinger Liquid.} 
\author {A. M. Tsvelik} 
\maketitle 
\begin {verse}
$Department~ of~ Physics,~ University~ of~ Oxford,~ 1~ Keble~ Road,$
\\
$Oxford,~OX1~ 3NP,~ UK$\\
cond-mat/9409027
\end{verse}
\begin{abstract}
\par
 We establish the equivalence between the problem of a potential
scatterer in a Luttinger liquid of spinless repulsive fermions and an
anisotropic Kondo model where the value of the impurity spin is
determined by the scaling dimension of the scattering potential.
The Bethe ansatz is used
to derive non-perturbative expressions for
the capacitance $C(V)$ and the conductance G(V) ($V$ is the voltage).
\end{abstract}

%\newpage
\sloppy
\par
%\section{Introduction}
 As Kane and Fisher have recently shown$^1$ a potential scatterer embedded
into a one-dimensional conductor acts at low energies as  a perfect
reflector if interactions between the electrons are predominantly
repulsive. As we shall see later, this effect can be described in the
framework of the anisotropic Kondo model where the value of the
impurity spin is determined by the scaling dimension of the
interaction. This equivalence is particularly striking because
from the  physics'
point of view these  two problems are  mirror images of each
other: in the
Kondo problem noninteracting  electrons become strongly  coupled
to the impurity because the latter has
internal degrees of freedom; in
the Kane-Fisher problem the impurity is just a potential scatterer and the
strong coupling develops due to  the electron-electron  interactions in
the bulk.

As in the Kondo effect, the reflection in a one-dimensional
Luttinger liquid is
perfect only at $T = 0, V = 0$ ($V$ is the voltage across the system);
at non-zero temperatures or voltages
there is a finite conductivity through the
impurity center: $G(\omega, V, T) \sim \max(\omega,T,V)^{2\lambda}$[1,2],
with  $\lambda > 1$.
 In this letter we use exact methods to
study electrostatic and transport properties of  an impurity center
in a one-dimensional  Luttinger liquid of spinless particles.
 As would be  expected the system behaves in a
very nonlinear fashion. We calculate the exact crossover functions at
T = 0 for the electric charge
$Q(V)$ and the current $I(V)$ through the barrier. It turns out
that in the limit $V \rightarrow 0$ (i) the capacitance $C_0 = \lim_{V
\rightarrow 0}Q/V$ is finite and
determined by the strength of the potential and by the parameters of the
Luttinger liquid, (ii) the conductance through the impurity center
decreases at small $V$ as a power law $G \sim V^{2\lambda}$.
The finiteness of the capacitance allows us to achieve good transport
through the impurity at
finite frequences. Since the real part of the resistance
of the barrier is given by
\begin{equation}
R = \frac{G}{[G^2 + (\omega C)^2]} \sim \frac{T^{2\lambda}}{\omega^2C^2}
\end{equation}
at sufficiently low temperatures $T < (\omega C)^{1/2\lambda}$
one enters into a regime of good conductance.

  Let us consider a model of spinless
fermions with a repulsive interaction in the presence of a single
potential scatterer placed at $x = 0$. In the continuous limit, where
only relevant interactions are kept, this model is described by the
following effective action:
\begin{equation}
S = \int \mbox{d}\tau \int_{- L}^L\mbox{d}x\left[
\frac{1}{2}(\partial_{\tau}\phi)^2 +
\frac{1}{2}(\partial_{x}\phi)^2 + M\cos(\beta\phi/\sqrt 2)\delta(x)\right]
\label{eq:act}
\end{equation}
This effective action can be further simplified by introducing
symmetric and antisymmetric phase combinations:
\begin{eqnarray}
\phi_{\pm}(x) = \frac{1}{\sqrt 2}\left(\phi(x) \pm \phi(-x)\right)
\end{eqnarray}
Then the nontrivial part of the action (\ref{eq:act}) contains only
$\phi_+$ and
acquires the following
form:
\begin{eqnarray}
S_+ = \int \mbox{d}\tau \int_0^{L}\mbox{d}x \left[
\frac{1}{2}(\partial_{\tau}\phi_+)^2 +
\frac{1}{2}(\partial_{x}\phi_+)^2 + M\cos(\beta\phi_+/2)\delta(x)\right]
\end{eqnarray}
This model is a particular case of the so-called boundary sine-Gordon
(BSG)
model solved exactly in Refs. 3, 4:
\begin{eqnarray}
S_{BSG} = \int \mbox{d}\tau \int_0^{L}\mbox{d}x [
\frac{1}{2}(\partial_{\tau}\phi)^2 +
\frac{1}{2}(\partial_{x}\phi)^2 + \nonumber\\
m\cos(\beta\phi) +
M\cos[\beta(\phi - \phi_0)/2]\delta(x)]
\end{eqnarray}

  The authors of Refs. 2, 3 have used the bootstrap procedure to
obtain the equations for
eigenvalues of momenta of physical particles, i.e.
excitations over the exact ground state. Their solution
is  expressed in terms of matrix elements of two-body
scattering matrices for true excitations. This form is not very
convenient for comparison
with exact solutions of other impurity problems (like Kondo
and Anderson models) where great experience has been accumulated
(see, for example, Refs. 5, 6 for a review).
To make contact with Kondo models easier we rewrite  the Bethe ansatz
equations in the equivalent form:

\begin{eqnarray}
\left\{\frac{\sinh[\pi\Delta(u_a + 1/g_1 - \mbox{i}/2)]\sinh[\pi\Delta(u_a -
1/g_1 - \mbox{i}/2)]}{\sinh[\pi\Delta(u_a + 1/g_1 +
\mbox{i}/2)]\sinh[\pi\Delta(u_a -
1/g_1 + \mbox{i}/2)]}\right\}^{N_0}\nonumber\\
\times\frac{\cosh[\pi\Delta(u_a - 1/g_2 +
\mbox{i}/2)]}{\cosh[\pi\Delta(u_a - 1/g_2 - \mbox{i}/2)]} =
\prod_{b = 1}^{M'}\frac{\sinh[\pi\Delta(u_a - u_b -
\mbox{i})]}{\sinh[\pi\Delta(u_a - u_b + \mbox{i})]} \label{eq:bethe}
\end{eqnarray}
Here $\Delta = \beta^2/8\pi$ is the conformal dimension of the
boundary term. The quantities $g_1, g_2$ are non-universal parameters
related to the bulk and
boundary masses $m$ and $M$ and the cut-off.
The integer number $N_0$ is proportional
to the length of the system $L$.
The quantities $u_a$ (rapidities) determine energy
eigenvalues of the system:

\begin{equation}
E = \sum_{a = 1}^{M'}\frac{1}{2\mbox{i}}
\ln\left\{\frac{\sinh[\pi\Delta(u_a + 1/g_1 - \mbox{i}/2)]\sinh[\pi\Delta(u_a -
1/g_1 - \mbox{i}/2)]}{\sinh[\pi\Delta(u_a + 1/g_1 +
\mbox{i}/2)]\sinh[\pi\Delta(u_a -
1/g_1 + \mbox{i}/2)]}\right\}
\end{equation}
In the ground state $M' = N_0/2 \rightarrow \infty$ and therefore the
rapidities $u_a$ do not represent physical excitations.

 In what follows we shall be interested only in the case where the
bulk mass is zero $m = 0$. This limit can be obtained by setting
\[
g_1, g_2 \rightarrow 0; \: \frac{N_0}{L}\exp[ - \frac{\pi\Delta}{1 -
\Delta}(1/g_2 - 1/g_1)] = const
\]
We restrict our attention to the case $\Delta = 1/(1 + \lambda)$ where
$\lambda > 1$ is an integer. Then
Eqs.(\ref{eq:bethe}) may be regarded as a result  of diagonalization
of the
following transfer matrix:
\begin{eqnarray}
T(u) =
\mbox{Tr}_0\left[R_{01}^{(1,1)}(u)R_{02}^{(1,1)}(u)...R_{0N}^{(1,1)}(u)R_{0,N
+ 1}^{(1,\lambda)}(u - 1/g)\right]\nonumber\\
E = \mbox{i}\frac{N_0}{L}\ln T(u = 0) \label{eq:prob}
\end{eqnarray}
where the trace is taken with respect to indices of the zeroeth
particle. The two-body scattering matrix $R^{(1,1)}$ describes a
scattering of two spin-1/2 particles; the matrix $R^{(1,\lambda)}$
describes a scattering of spin-1/2 and spin $S = \lambda/2$
particles. These matrices are solutions of the Yang-Baxter equation
and have the following form:
\begin{eqnarray}
R_{0j}^{(1,1)}(u) = \sum_{i =
0}^3w_i(u)\sigma^i_0\otimes\sigma_j^i\nonumber\\
R_{0,N + 1}^{(1,2S)}(u) = \sum_{i =
0}^3w_i(u)\sigma^i_0\otimes I^i\nonumber\\
w_0(u) = \frac{\sinh(u +
\mbox{i}\pi\Delta/2)}{\mbox{i}\sin(\pi\Delta/2)}, \: w_1 = w_2 = 1, \:
w_3(u) = \frac{\cosh(u +
\mbox{i}\pi\Delta/2)}{\cos(\pi\Delta/2)}\nonumber\\
I^0 = \frac{\cos(\pi\Delta S^z)}{2\cos(\pi\Delta/2)}, \: I^3 =
\frac{\sin(\pi\Delta S^z)}{2\sin(\pi\Delta/2)}, \nonumber\\
I^+ = I^1 + \mbox{i}I^2 = S^+f(S^z), \: I^+ = I^1 - \mbox{i}I^2 =
f(S^z)S^-\nonumber\\
f(S^z) = \frac{1}{\sin(\pi\Delta)}\left[\frac{\sin\pi\Delta(S + 1 +
S^z)\sin\pi\Delta(S - S^z)}{(S + 1 + S^z)(S - S^z)}\right]^{1/2}
\end{eqnarray}
where in our case $S = \frac{1}{2}\lambda$.
The matrix $R^{(1,2S)}$ with general $S$
was  found by Fateev (unpublished) and Kulish and
Reshetikhin [7]. The problem (\ref{eq:prob}) arises as a part of the Bethe
ansatz
solution of
the anisotropic Kondo model  with  spin-1/2 electrons scattering  on the
impurity with spin $S = \frac{1}{2}\lambda$:
\begin{equation}
H = \int \mbox{d}x\left\{ -
\mbox{i}\psi^+_{\alpha}\frac{\mbox{d}}{\mbox{d}x}\psi_{\alpha} +
\delta(x)\psi^+_{\alpha}\left[I_{\parallel}\sigma^z S^z +
I_{\perp}(\sigma^xS^x +
\sigma^yS^y)\right]_{\alpha\beta}\psi_{\beta}\right\} \label{eq:kondo}
\end{equation}
where the constants  $I_{\parallel}, \: I_{\perp}$
($I_{\parallel} > I_{\perp}$) are related
to $\Delta$ and $g^{-1} = g_2^{-1} - g_1^{-1}$ in a non-universal way. We
regard the  equivalence between the boundary sine-Gordon model (5)
with the zero bulk mass $m = 0$ and the
anisotropic Kondo model (\ref{eq:kondo}) as rather unexpected.
This equivalence may give  a clue for constructing
other exactly solvable boundary
problems.

 In the thermodynamic limit solutions of
Eqs.(\ref{eq:bethe}) consist of solutions with a constant imaginary
part (``shifted'' rapidities)
\begin{equation}
u_a = \frac{\mbox{i}}{2\Delta} + u^{(0)}_a
\end{equation}
and ``strings''
\begin{equation}
u_a = \frac{\mbox{i}}{2}(n + 1 - 2j) + u^{(n)}_a, j = 1, 2, ... n; n \leq
\lambda
\end{equation}
The ground state is filled with shifted rapidities. To describe
excitations in the thermodynamic limit we introduce the distribution
functions  of holes in the ground state $\rho_0(u)$ (they correspond
to solitons) and shifted
rapidities $\tilde\rho_0(u)$ as well as the distribution functions
$\rho_n(u)$ of
real parts of strings $u^{(n)}$  and their holes
$\tilde\rho_n(u) (n = 1, 2, ... \lambda)$ (note the difference in the
notations between solitons and other particles).
In the sine-Gordon model notations
the distributions with $0 < n < \lambda$ correspond to  breathers
and $\rho_{\lambda}$ describes antisolitons.
The equations for these distribution functions
have the following form:
\begin{eqnarray}
\tilde\rho_n(u)  + B_{nm}\star\rho_m(u) =
A_n\exp(- \pi u/2) + \frac{1}{L}f_n(u -
1/g) \label{eq:cont}
\end{eqnarray}
where the star denotes  a convolution
\[
f\star g(u) = \int \mbox{d}u' f(u - u')g(u')
\]
 and the Fourier images of the kernels are
\begin{eqnarray}
B_{00}(x) = B_{\lambda\lambda}(x) = B_{0\lambda}(x) + 1 =
\frac{\sinh[(\lambda^{-1} + 1)x]}
{2\sinh[\lambda^{-1}x]\cosh x}, \nonumber\\
B_{nm}(x) = \frac{2\coth(\lambda^{-1}x)\cosh[(1 -
n\lambda^{-1})x]\sinh[m\lambda^{-1}x]}{\cosh x}, \lambda >
n \geq m; \nonumber\\
B_{0n}(x)  = \frac{2\coth(\lambda^{-1}x)\sinh[(1 +
\lambda^{-1} -
n\lambda^{-1})x]}{\cosh x}; \nonumber\\
f_0(x) =  f_{\lambda}(x) = \frac{\tanh x}{2\sinh(\lambda^{-1}x)};  f_n(x) =
\frac{\sinh(\lambda^{-1}nx)}
{\sinh(\lambda^{-1}x)\cosh x} (0 < n < \lambda^{-1})\nonumber\\
A_n = 2\sin(\pi n/\lambda)A_0  \mbox{ at $0 < n < \lambda^{-1}$}, A_{\lambda} =
A_0.
\end{eqnarray}
Notice that the present expressions for
the kernels $B_{0n}, \: f_0,  \: f_{\lambda}$ are different from the ones
given in Ref. 3.

  The original Bethe ansatz equations  (\ref{eq:bethe}) depend
explicitly on the number $M'$.
The quantity which is proportional to $L$  in the thermodynamic limit {\it and
vanishes in the ground state} is
\begin{equation}
Q = \left(N_0/2 - M' + \frac{\lambda}{2}\right) = \frac{(\lambda + 1)}{2}L\int
\mbox{d}u [\rho_0(u) - \rho_{\lambda}(u)]
\end{equation}
In the conventional  sine-Gordon model $Q$ is proportional to  the
topological charge, the latter being
the number of solitons minus the number of
antisolitons. In the boundary problem the topological charge is not
conserved because solitons can transform into
antisolitons and vice versa when scattering on the boundary.
Therefore it may appear difficult to find a physical meaning for
 a  field
conjugated to $\theta$ in the boundary problem.
Here the equivalence with the Kondo model comes to our help. For the
Kondo model we know that $Q$ represents the $z$-projection of
the total spin, i.e.  it includes spins of the electrons and the impurity.
We also know that if  $T_K << N_0/L$ (the latter quantity is the bandwidth)
the relative contribution of the impurity to the magnetic
susceptibility is much greater than that of the electrons. For this
reason in the Kondo problem one can  consider a magnetic field as
acting on the impurity spin only. Acting in a similar way one can
consider the impurity part of $Q$ as the electric charge associated
with the boundary and consider it as a quantity conjugated to the
potential difference through the barrier $eV$.

 At T = 0 the only nonvanishing densities are $\rho_0$ and
$\tilde\rho_0$, the former being non-zero only in the interval $(b,
\infty)$, where $b$ is determined by  the bulk charge susceptibility
and $V$. As
follows from Eqs.(\ref{eq:cont}) these
two densities satisfy  the following integral equation:
\begin{equation}
\tilde\rho_0(u) +  \int_b^{\infty} \mbox{d}u' B_{00}(u - u')\rho_0(u') = A_0
\exp( - \pi u/2) + \frac{1}{L}f_0(u - 1/g) \label{eq:hopf}
\end{equation}
The limit $b$ is determined by the charge-voltage
relationship in the bulk:
\begin{equation}
eV\chi = \frac{\lambda + 1}{2}\int_b^{\infty} \mbox{d}u
\rho_0^{bulk}(u) \label{eq:bulk}
\end{equation}
where $\chi$ is the charge susceptibility of the Luttinger liquid.
Equation (\ref{eq:hopf})
is a standard Wiener-Hopf equation of the type discussed in great
detail in the literature, for example, in Refs. 5,6.
As we have said, the electrical charge on the impurity is  related to
the impurity part of $Q$:
\begin{eqnarray}
\frac{1}{e}Q^{imp} = \frac{\lambda + 1}{2}\int_b^{\infty} \mbox{d}u
\rho_0^{imp}(u) = \nonumber\\
\frac{1}{4\pi}\int\mbox{d}x \frac{\exp[-
\frac{2\mbox{i}x}{\pi}\ln(T_K/V)]}{x -
\mbox{i}0}\frac{1}{G^{(-)}(x)}\frac{\sinh
x}{\sinh(\lambda^{-1}x)\cosh x}
\end{eqnarray}
where $T_K \sim A_0\exp(- \pi /2g) \sim M^{1+ 1/\lambda}$
is the characteristic energy scale of the impurity and $G^{(\pm)}(x)$ are
functions analytical in the upper (lower)
half-plane of $x$:
\begin{equation}
G^{(+)}(-x) = G^{(-)}(x) =  \frac{\Gamma(1 +
\frac{\mbox{i}x}{\pi\lambda})\Gamma(\frac{1}{2} +
\frac{\mbox{i}x}{\pi})}{\Gamma(1 + \frac{\mbox{i}x(\lambda +
1)}{\pi\lambda})\Gamma(1/2)}
\end{equation}
The result for $Q^{imp}$
can be written as two different  series expansions
\begin{eqnarray}
\frac{1}{e}Q^{imp} = \nonumber\\
\frac{1}{\pi}\sum_{n = 0}^{\infty}\frac{\Gamma(1/2)\Gamma[1
+ (1 + \lambda^{-1})(n + 1/2)]}{\Gamma[1
+ \lambda^{-1}(n + 1/2)]n!}\frac{1}{(2n + 1)\sin\left(\frac{\pi(2n +
1)}{2\lambda}\right)}\left(\frac{eV}{T_K}\right)^{(2n + 1)} \mbox{ at $V <
T_K$}\label{eq:Q}\\
\frac{1}{e}Q^{imp} = \nonumber\\
\frac{\lambda}{2} - \frac{\lambda + 1}{2\pi}\sum_{n =
1}^{\infty}\sin\left(\frac{\pi n}{\lambda + 1}\right)\frac{\Gamma(1 +
\frac{n}{1 + \lambda})\Gamma[1/2 + n(1 -
\frac{1}{1 + \lambda})]}{n! n}\left(\frac{T_K}{eV}\right)^{\frac{2n\lambda}{1 +
\lambda}}
 \mbox{at $V > T_K$}
\end{eqnarray}
It follows from Eq.(\ref{eq:Q}) that the capacitance at $V = 0$ is
finite:
\begin{equation}
C_0 = \frac{e^2}{T_K}\frac{\Gamma(3/2 + 1/2\lambda)\Gamma(1/2)}
{\Gamma(1 + 1/2\lambda)\sin(\pi/2\lambda)}
\end{equation}

 Now we calculate the current through the barrier using the expression
derived  in
Ref. 2 for a similar problem.
Adjusting the notations in Eq.(14) of Ref. 2 to those used
in this paper, we get
\begin{equation}
I(V) = ev_F\int \mbox{d}u |S_{++}(u - 1/g)|^2(\rho_0^{bulk}(u) -
\rho_{\lambda}^{bulk}(u)) \label{eq:cur}
\end{equation}
where $S_{++}(u)$ is the scattering amplitude of a soliton by the
impurity and $v_F$ is the Fermi velocity. This amplitude was found in Refs. 3,
4. In our notations we
have
\begin{equation}
|S_{++}(u)|^2 = \frac{1}{e^{\pi\lambda u} + 1}
\end{equation}
Substituting into (\ref{eq:cur})
the solution of the Wiener-Hopf equation (\ref{eq:hopf})
for the bulk density $\rho_0^{bulk}(u)$ and using condition
(\ref{eq:bulk}) we get
\begin{eqnarray}
I(V) = \frac{\mbox{i}e^2V(\chi v_F)}{\lambda(\lambda + 1)}\int \mbox{d}x
\frac{1}{\lambda(\lambda +
1)}\frac{\exp[2\mbox{i}x/\pi\ln(eV/T_K)]}{\lambda\sinh[\lambda^{-1}(x
- \mbox{i}0)]}\frac{1}{(\pi + 2\mbox{i}x)G^{(-)}(x)}
%\int_0^{\infty}\frac{du}{e^{\pi\lambda
%u}(T_K/eV)^{2\lambda} + 1}F(u)
\end{eqnarray}
%where $F(u)$ is the universal function
%$F(u) \sim  \rho_0^{bulk}(u - b)/eV$ and the Fourier image of
%this function is given by
%\begin{equation}
%F(x) = \frac{G^{+}(0)}{G^{+}(x)}\frac{\pi}{\pi - 2ix}
%\end{equation}
At small $eV << T_K$ we find
\begin{equation}
G = \frac{I}{V} = \frac{e^2}{12\pi}\frac{(\lambda +
1)\Gamma(1/2)\Gamma(2 + \lambda)}{\Gamma(1/2 + \lambda)}(eV/T_K)^{2\lambda}
\label{eq:G}
\end{equation}
At large $V$ the conductance  achieves its maximum value $G_{max} =
2e^2(v_F\chi)/(\lambda + 1)$. For the model of spinless fermions the quantity
$(v_F\chi)$ is related to the conformal dimension of the perturbation:
\begin{equation}
(v_F\chi) = \frac{(1 + \lambda)}{4\pi}
\end{equation}
Therefore the maximal conductance is equal to
\begin{equation}
G_{max} = \frac{e^2v_F}{2\pi}
\end{equation}
Comparing Eq.(\ref{eq:G}) and  Eq.(\ref{eq:Q}) we find
that a small
electric charge disappears as
\begin{equation}
 Q \sim e(tT_K)^{1/(2\lambda - 1)}
\end{equation}
For large charges the exponent changes:
\begin{equation}
e\lambda/2 - Q = (tT_K)^{\frac{2\lambda}{3\lambda + 1}} \label{eq:ch}
\end{equation}


\begin{thebibliography}{99}
\bibitem{1} C. L. Kane and M. P. A. Fisher, Phys. Rev.  {\bf B46},
15233(1992).
\bibitem{2} P. Fendley, A. W. W. Ludwig and H. Saleur,
cond-mat/9408068.
\bibitem{3}
S. Ghoshal and A. B. Zamolodchikov, Int. J. Mod. Phys.
{\bf A9}, 3841 (1994).
\bibitem{4} P. Fendley, H. Saleur and N. P. Warner, to appear in Nucl. Phys. B,
hep-th/9406125.
\bibitem{5} N. Andrei, K. Furuya and J. Lowenstein, Rev. Mod. Phys.
{\bf 55}, 331 (1983).
\bibitem{6} A. M. Tsvelik and P. B. Wiegmann, Adv. Phys. {\bf 32}, 453
(1983).
\bibitem{7} P. P. Kulish and N. Yu. Reshetikhin,
Zap. Nauch. Semin. LOMI {\bf 101}, 101 (1981). The results of this
paper are cited in H. M. Babujian and A. M. Tsvelik, Nucl. Phys. B {\bf
263}, 24 (1986).




\end{thebibliography}
\end{document}